\documentclass[mathleft
]{an}
\usepackage{graphicx}
\usepackage{times}

\usepackage{amssymb}                    
\usepackage{color}                       
\usepackage{url}                         
\usepackage[utf8]{inputenc}
\usepackage{multirow}
\usepackage{bbm}
\usepackage{txfonts}

\begin{document}

\Pagespan{789}{}
\Yearpublication{2006}%
\Yearsubmission{2005}%
\Month{11}%
\Volume{999}%
\Issue{88}%

\title{Solar bursts as can be observed from the lunar farside with a single antenna at very low frequencies}

\author{A.A. Stanislavsky\inst{1,2}\thanks{\email{a.a.stanislavsky@rian.kharkov.ua}\newline}
\and A.A. Konovalenko\inst{1} \and S.N. Yerin\inst{1,2} \and I.N. Bubnov\inst{1} \and V.V. Zakharenko\inst{1,2} \and Yu.G. Shkuratov\inst{1,2} \and P.L. Tokarsky\inst{1} \and Ya.S. Yatskiv\inst{3} \and A.I. Brazhenko\inst{4} \and A.V. Frantsuzenko\inst{4} \and V.V. Dorovskyy\inst{1} \and H.O. Rucker\inst{5} \and Ph. Zarka\inst{6}
}
\titlerunning{Solar bursts as can be observed from the lunar farside with a single antenna $\dots$}
\authorrunning{A.A. Stanislavsky et al.}
\institute{Institute of Radio Astronomy, National Academy of Sciences of Ukraine, 4, Mystetstv St., 61002, Kharkiv, Ukraine
\and V.N. Karazin Kharkiv National University, Svobody Sq., 4, 61022 Kharkiv, Ukraine
\and Main Astronomical Observatory, National Academy of Sciences of Ukraine, 27, Ak. Zabolotnoho St., 03143 Kyiv, Ukraine
\and Institute of Geophysics, National Academy of Sciences of Ukraine, 27/29, Myasoedova St., 36014 Poltava, Ukraine
\and Commission for Astronomy, Austrian Academy of Sciences, 6, Schmiedlstrasse, 8042 Graz, Austria
\and LESIA \& USN, Observatoire de Paris, CNRS, PSL/SU/UPMC/UPD/SPC, Place J. Janssen, 92195 Meudon, France
}

\received{data} \accepted{data} \publonline{later}

\keywords{Sun: corona -- Sun: radio radiation -- methods:
observational -- instrumentation: miscellaneous -- telescopes}

\abstract{%
Earth-based observations are complicated by the opacity of Earth's ionosphere at very low frequencies and strong man-made radio frequency interference. This explains long standing interest in building a low frequency radio telescope on the farside of the Moon. Experience from ground-based observations near the ionospheric cutoff in dealing with the interference, ionosphere, and wide--field imaging/dynamic range problems provides crucial information for future radio--astronomic experiments on the Moon. In this purpose we observed non--intensive solar bursts on the example of solar drift pairs (DP) at decameter--meter wavelengths with large and small arrays as well as by a single crossed active dipole. We used the large Ukrainian radio telescope UTR--2, the URAN--2 array, a subarray of the Giant Ukrainian radio telescope (GURT) and a single crossed active dipole to get the spectral properties of radio bursts at the frequency range of 8--80 MHz during solar observations on July 12, 2017. Statistical analysis of upper and lower frequencies, at which DPs are recorded, shows that the occurrence of forward DPs is more preferable at lower frequencies of the decameter range of observations in comparison with reverse DPs generated more likely at meter wavelengths. We conclude that DPs can be detected not only by antenna arrays, but even by a single crossed active dipole. Thus the latter antenna has a good potential for future low--frequency radio telescopes on the Moon.
}

\maketitle

\section{Introduction}

Low frequency radio astronomy on the lunar farside has several important advantages over ground-based studies (Jester and Falcke 2009, Mimoun et al. 2012 and references therein). The main ones are no ionosphere, impenetrable for radio emission with frequencies lower than about 10--15 MHz and causing strong signal distortions at low frequencies, and the lack in the frequency range below 30 MHz of a strong electromagnetic pollution due to man--made activities (broadcasting, telecommunications) and natural lightning. However, the research of this part of the radio spectrum is of great interest. It is necessary to mention here such phenomena as coronal mass ejections (CMEs), solar radio bursts, planetary radio emission, extrasolar planetary radio bursts and many others (Klein-Wolt et al. 2012, Zarka et al. 2012). Opening a new radio--astronomic window, one may anticipate interesting discoveries, both predictable and unexpected ones. Preliminary studies in these areas seem to be very useful. In this paper we focus on solar radio bursts with a moderate flux, which include solar drift pairs (DPs). Many cosmic radio sources, in fact, have strong emissions at the wavelengths, but many ones are relatively weak. This imposes certain requirements in building a low frequency radio telescope on the farside of the Moon. Before deploying and maintaining any antenna on the lunar surface, it should be chosen its design, preferring either an array of simple antennas or a high--quality single antenna. Using recent ground--based observations, we are going to compare the observational results, obtained in both cases, for relatively weak sources (such as DP bursts, for example). In addition, the reasons for why these events can be expected at such low frequencies are discussed below.

\begin{figure}
\centering
\resizebox{1.\columnwidth}{!}{%
  \includegraphics{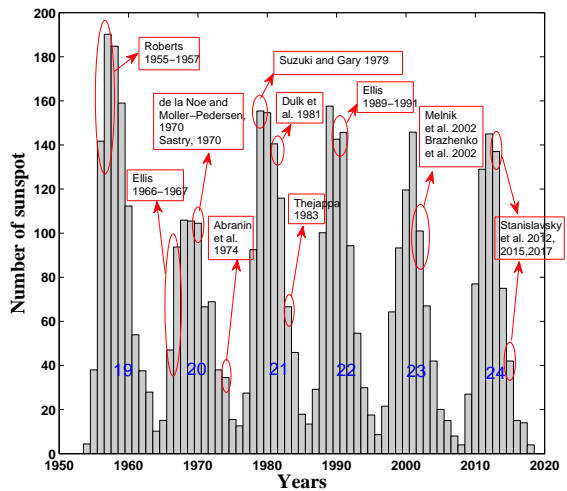}
} \caption{Overview of DPs radio observations during recent solar cycles (see more  details in the relevant references of this paper).} \label{sunspot}
\end{figure}

Starting with the pioneering work at the dawn of radio astronomy (Roberts 1958), the studies of DPs have already been continued for six decades up to now. Figure~\ref{sunspot} briefly shows their chronicle. The type of solar bursts was observed by different instruments during both growth and decline of solar activity as well as near maxima. Solar DPs have a simple shape on dynamic spectra of radio records recognized as two narrow components separated in time; often the second component is a repetition of the first. They are detected by ground-based instruments at decameter-meter wavelengths, but each individual DP occupies only a limited bandwidth in the frequency range, they drifting whether from higher frequencies to lower ones (forward DPs, briefly FDPs) or visa versa (reverse DPs, RDPs). No each storm generates a DP during the solar storms of type III bursts. The type III bursts are thought to be produced by the accelerated electrons propagating along open magnetic field lines in the solar corona. The role of electron beams in the generation of DPs remains unclear yet. Moreover, the frequency--time features of DPs, such as variety of frequency drift rates from one pair to another, appearance of vertical DPs and others, are also unexplained.

In the paper (Melnik et al. 2005), based on solar decameter observations in July of 2002, the total frequency bandwidths occupied by DPs were measured, but the radio records in the main were made by the analogous multichannel receiver (10--30 MHz), tuned to selected 60 frequencies with the frequency bandwidth 3 kHz in each frequency channel. The frequency gaps between neighbor frequency channels in the 60--channel spectrometer ranged from 110 kHz to 1.4 MHz (most of them had 300 kHz in frequency spacing) in dependence of radio interference environment at 10--30 MHz. Unfortunately, the interference ``weather'' at decameter wavelengths is complex and changeable, especially in daytime. Therefore, the receivers of this spectrometer slightly were tunable from time to time. As a result, the frequency--bandwidth histograms for FDPs and RDPs were distorted because not all DPs have been identified, and the long DPs observed were not taken into account at all. For extending and updating our knowledge about the DP distributions over frequency, the radio observations of DPs are continued now (Litvinenko et al. 2016).  In the summer of 2015 their studies were performed in the continuous frequency band of 8--32 MHz with high resolution in time and frequency (see Volvach et al. 2016, Stanislavsky et al. 2017a, 2017b). The observations were carried out with the Ukrainian T--shape radio telescope (UTR--2) located near Kharkiv. They clearly indicate a difference in behavior of FDPs and RDPs in histograms of starting and ending frequencies. In our paper the results of new DP observations are presented using the Ukrainian low--frequency radio instruments (Konovalenko et al. 2016), that have already delivered important astrophysical results.

\section{The radio instruments}

During the observational session in 10--16 July of 2017, the solar observations were performed with several radio instruments differing in their characteristics, which we will present briefly in this section.

\subsection{UTR-2}

The UTR--2 radio telescope, located at coordinates 36$^\circ$ 56$^\prime$ 29$^{\prime\prime}$ E and +49$^\circ$ 38$^\prime$ 10$^{\prime\prime}$ N, has a T-shape (approximately 2 km $\times$ 1 km) and splits into 12 subarrays (sections) that form three arms (North, South and West). The instrument is characterized by a number of unique features: large effective area (about 150 000 m$^2$), electronic control of the beam steering within a wide range of the two sky coordinates ($\pm$ 70$^\circ$ from zenith in N--S and E--W directions), high filling factor, high directivity (the width of the main beam is about 0.5$^\circ$), low level of side lobes, wide frequency range (8--32 MHz), simultaneous multibeam capability, high dynamic range, interference immunity, reliability and others. This makes it very powerful and flexible for overcoming obstacles that hinder ground-based observations at decameter wavelengths (Konovalenko et al. 2016). There is no need to use the UTR--2 in the full antenna configuration for the solar observations, and we used four sections of the north--south array. In this case the beam pattern size of $1^\circ\times 15^\circ$ at 25 MHz covered the solar corona from which solar bursts were expected.

\subsection{URAN-2}

To operate jointly with the UTR--2 as a VLBI network or to record signals independently at spaced arrays, a system of decameter interferometers (URAN) was built in Ukraine (Konovalenko et al. 2016). Unlike the UTR--2 providing measurements in one linear polarization, the URAN arrays have dual-polarized antennas. The system consists of four radio telescopes of smaller sizes. The URAN--2 is one of them, located in Stepanivka village near Poltava. The rectangular array contains 512 crossed dipoles in a 238 $\times$ 118 m. The distance between UTR--2 and URAN--2 is about 153 km. After the full UTR--2 the URAN--2 radio telescope is the second in size, its effective area being 28 000 m$^2$. As applied to the given measurements, the URAN--2 array is comparable to four sections we used in the UTR--2.

\subsection{GURT}

The GURT system develops the UTR--2 in terms of spatial dimensions and frequency range located nearby it (Konovalenko et al. 2016, Stanislavsky et al. 2014). The basic part of this telescope is a square regular subarray using active dipole techniques. Each subarray is made of 25 crossed dipoles (5 $\times$ 5). The distance between the crossed dipole antenna elements is 3.75 m, and the suspension height is 1.6 m. The subarray design provides a wide frequency coverage from 8 to 80 MHz, high sensitivity (the galactic background level exceeds their self--noise by more than 7 dB), and high RFI immunity due to high dynamic range of the dipole amplifier (input IP3 is 30 dBm). The effective area at the central frequency, confirmed by computer simulations and direct measurements, is about 350 m$^2$. The apparent advantages of the GURT over the UTR--2 are the ability to measure full polarization characteristics of the received radio waves and to track the source down to the altitude of a few arc degrees above the horizon. The GURT subarray is inferior to each UTR--2 section in sensitivity in the same frequency range, but the former has more than two times wider frequency band for observations.

\subsection{Crossed active dipole}

The crossed active dipole is a cornerstone of GURT subarrays. There is no doubt that the better the elemental antenna is adapted for radio astronomy, the better the array made of them. It was given great attention in the development of the GURT system. In this connection it should be noticed the role of active antennas. They have very useful advantages in comparison with passive ones. In particular, below 30 MHz, where the external (Galactic) noise exceeds the internal one considerably, shortening the radiator length of a tuned antenna does not affect the signal--to--noise ratio at the antenna output, but shortening the radiator dramatically changes its input impedance, and therefore the preamplifier matches the dipole impedance back to the cable one. Thereby the length of a short-wave antenna can be reduced noticeably. Consequently, the combination of the dipole and preamplifier suggests the maximally possible ratio between the antenna temperature caused by the Galactic noise $T_{sky}$ and that of the preamplifier noise, $T_{pre}$, namely $10\lg\,(T_{sky}/T_{pre}) \approx 10$ dB from 10 MHz to 70 MHz (Tokarsky et al. 2015). Each identical antenna of the GURT subarray is made of two crossed dipoles with stand-alone terminals to receive radiation of two orthogonal polarizations independently.  Each arm of a dipole is formed by three tubes that depart from the input terminal. Two of them form a triangle with the rounded vertexes, and the third tube goes along the triangle median line, that has sharp bend at 45$^\circ$. They are made of polyethylene--coated thin--walled copper tubes with 12 mm diameter.  All tubes are fastened by the standard fitting and in addition are bridged with copper wires. The total length of the dipole along the arm median is 2.8 m, and maximum width is 0.8 m. Both dipoles are mounted at a height of 1.6 m above the ground on the vertical rack made of a steel tube in diameter of 60 mm with the upper dielectric nozzle. In contrast to the LOFAR and LWA projects, we do not accomplish any special metallization of the ground, because it has a high conductivity at the place of location of the GURT antenna array (Tokarsky et al. 2017a). Note also that such a crossed active dipole antenna is more sensitive than each dipole of the UTR--2 array, not to mention the frequency range and polarization capabilities.

\subsection{Wideband digital receivers}

Low--frequency radio telescopes deal with large variations of received signals, including the appreciable Galactic--background radio emission and powerful radio frequency interference (RFI), particularly from broadcasting stations. This explains the application of a 16--bit ADC in digital receivers used for radio astronomy purposes in low frequencies. Operating in 8--32 MHz (UTR--2, URAN--2), the first own wideband digital receiver (named DSPZ from ``Digital Spectro--Polarimeter, type Z'') was developed in Institute of Radio Astronomy, Kharkiv in 2006. It has a sampling frequency at least twice larger than the maximum frequency of operation of the radio telescopes, performs on--line Fast Fourier Transform (FFT) or records raw waveforms. The UTR--2 is equipped with five sets of the two--channel receivers, one for each of the five separate and simultaneous telescope beams. The same type of the receiver is used for the URAN--2. Their main parameters are listed in the paper of Ryabov et al. (2010). The ADR (Advanced Digital Receiver) developed for GURT subarrays has a higher sampling frequency and thus a broader spectral range of operation. The design of this receiver was built on all the technical advances from the DSPZ, but also included a number of new capabilities (Zakharenko et al. 2016). The latter receivers cover the full GURT radio telescope band with good frequency-time resolutions and a high dynamic range.

\subsection{Sensitivity of one GURT active dipole}

\begin{figure}
\centering
\resizebox{1.\columnwidth}{!}{%
  \includegraphics{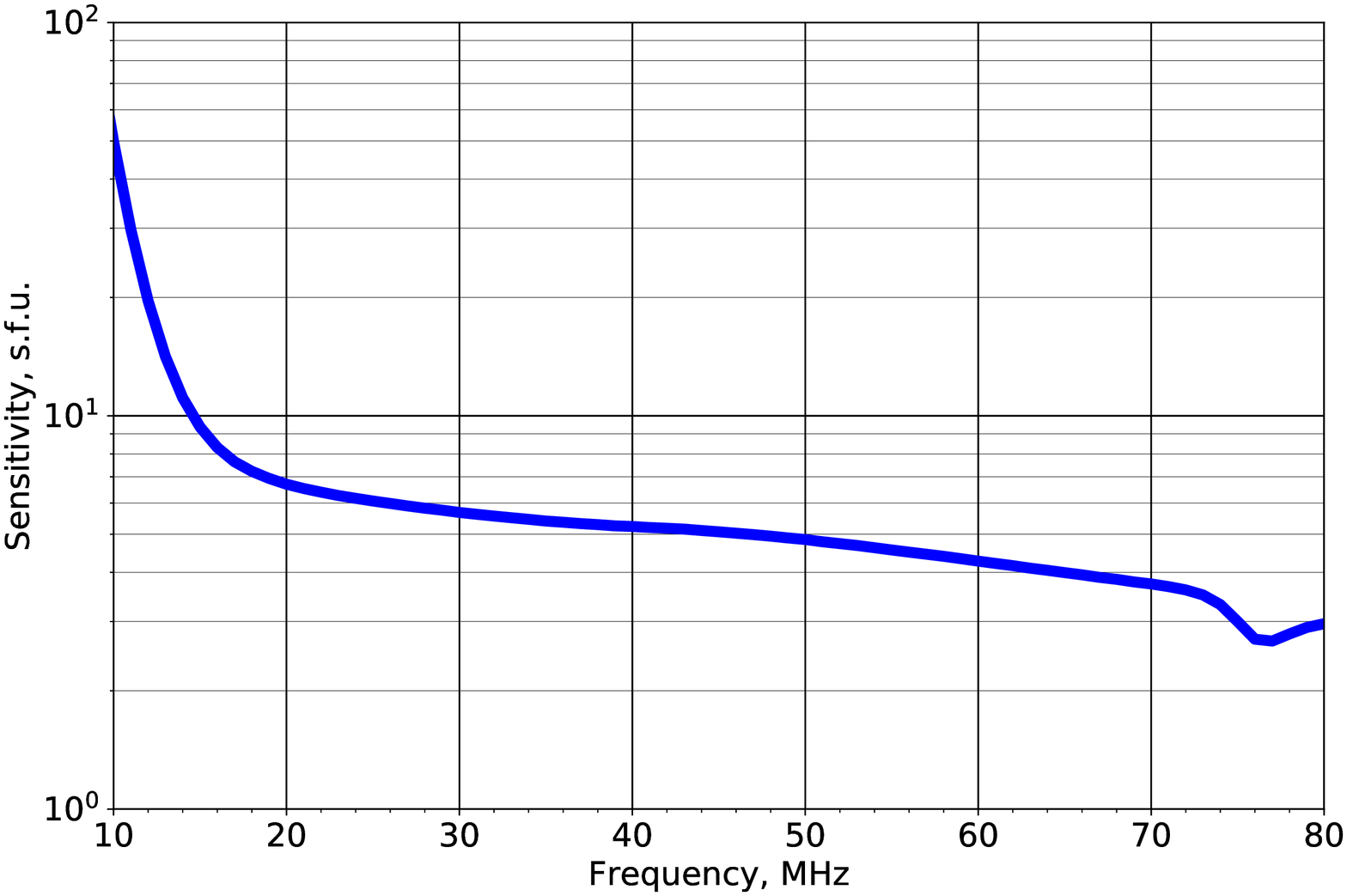}
} \caption{Sensitivity of the GURT active antenna in dependence of frequency.} \label{sensitivity}
\end{figure}

The performance of each radio telescope is characterized its sensitivity. Let us find it for one GURT active antenna element. From the well-known book of Kraus (1966) the general formula, describing the radio telescope sensitivity, is
\begin{displaymath}
\Delta S_{min}=K_{rec}\frac{2k_B T_{sys}}{A_e\sqrt{\Delta t\,\Delta f}}\,,
\end{displaymath}
where $\Delta S_{min}$ is the fluctuation sensitivity of the radio telescope (the minimal signal that can be detected over the background of fluctuation noise), $K_{rec}$ the coefficient taking into account the type of a receiver (in our case the receiver does not use any signal modulation, i.\,e. $K_{rec}=1$), $k_B$ the Boltzmann constant, $T_{sys}$ the system noise temperature, $A_e$ the effective area of antenna, $\Delta t$ the integration time after detector (temporal resolution), $\Delta f$ the signal frequency range before detector (frequency resolution). The dimensionless coefficient $\sqrt{\Delta t\,\Delta f}$ is also called the radiometric gain. To estimate the sensitivity of the GURT dipole – ADR receiver system, we accurately took into account internal and external noises of the system. Based on EM (electromagnetic) simulations that were verified with real noise measurements, the system equivalent flux density (SEFD) of the GURT antenna element was found (Tokarsky et al. 2017b). Note that SEFD accounts for both system noise and antenna effective area rather than the parameters of observations. The fluctuation sensitivity and SEFD are related as follows
\begin{displaymath}
\Delta S_{min}=\frac{SEFD}{\sqrt{\Delta t\,\Delta f}}\,.
\end{displaymath}
Using Fig.11 of Tokarsky et al. (2017b) that shows SEFD of the GURT antenna element for real noise parameters and for a noiseless system, we can easily determine the fluctuation sensitivity, which is drawn in Figure~\ref{sensitivity}. The sensitivity was calculated for typical parameters of solar observations: $\Delta t$ = 100 ms and $\Delta f$ = 9.765 kHz. At the central frequency (40 MHz) of the GURT operational range it results in 5 s.f.u. (1 s.f.u. = 10$^{-22}$ W m$^{-2}$ Hz$^{-1}$). Thus, the detection capability of the GURT antenna element, as applied to the solar bursts with a moderate flux, is quite encouraging.

\section{Observations}\label{observations}

On 10--16 July of 2017, according to the space--based observations with STEREO, GOES and SOHO, solar activity varied with time, from a low level on 11 July and moderate on 12 July to high on 14 July, again moderate on 15--16 July and low later. The most of solar events were accompanied with the active region NOAA AR 12665. Its evolution is presented in Table~\ref{ar12665}. The region moved along a solar disk with angular velocity 0.238 radians per day. It was the first sunspot to appear after the Sun was spotless for two days. During its 13--day trip across the face of the Sun, the active region produced several solar flares, a coronal mass ejection and a solar energetic particle event (see https://www.solarmonitor.org). Solar wind speed at SOHO ranged between 387 and 472 km/s. The dark core of the sunspot was larger than the Earth. After 12 July the active region decayed slowly and quietly. It is interesting that during the observations of DPs on 12 July of 2017 the active region was a bipolar sunspot group. A similar sunspot group was associated with DPs observed in 2015 (Stanislavsky et al. 2017a).

\begin{table}
\caption{Several features of AR 12665 (see all the notations in https://www.spaceweatherlive.com/en/solar-activity/region/12665). The third column is expressed in ``millionths of a solar hemisphere'' (MH).}
\label{ar12665}
\small
\resizebox{1.\columnwidth}{!}{
\begin{tabular}{c|ccccc}
\hline
Date    &      Sunspot    &   Size   &    Class     &    Class   & Location \\
        &      Number     &          &    Magn.     &    Spot    &          \\
\hline
2017-Jul-06	&	1      	&	70	 &	$\alpha$  &   HSX   &  S05E78  \\
2017-Jul-07	&	6      	&	140	 &	$\beta$   &   DAI	&  S06E64  \\
2017-Jul-08	&	15      &	310	 &	$\beta$   &   EKC	&  S06E52  \\
2017-Jul-09	&	14      &	480	 &	$\beta$-$\gamma$-$\delta$   &  EKC	& S07E38 \\
2017-Jul-10	&	21      &	710  &	$\beta$-$\gamma$   &  EKC	& S06E24  \\
2017-Jul-11	&	22      &	690  &	$\beta$-$\gamma$   &  EKC	& S06E11  \\
2017-Jul-12	&	17      &	620  &	$\beta$-$\gamma$   &  EKC	& S06W03  \\
2017-Jul-13	&	17      &	570  &	$\beta$   &  EHI	&  S06W17  \\
2017-Jul-14	&	26      &	440  &	$\beta$   &  EHI	&  S07W30  \\
2017-Jul-15	&	18      &	460  &	$\beta$   &  EKC	&  S06W43  \\
2017-Jul-16	&	12      &	340  &	$\beta$   &  DKI	&  S05W57  \\
2017-Jul-17	&	5       &	380  &	$\beta$   &  DKO	&  S06W70  \\
2017-Jul-18	&	3       &	450  &	$\beta$   &  CHO	&  S06W86  \\
2017-Jul-19	&	3       &	450  &	$\beta$   &  CHO	&  S06W0  \\
\hline
\end{tabular}
}
\end{table}

\begin{figure}
\centering
\resizebox{1.\columnwidth}{!}{%
  \includegraphics{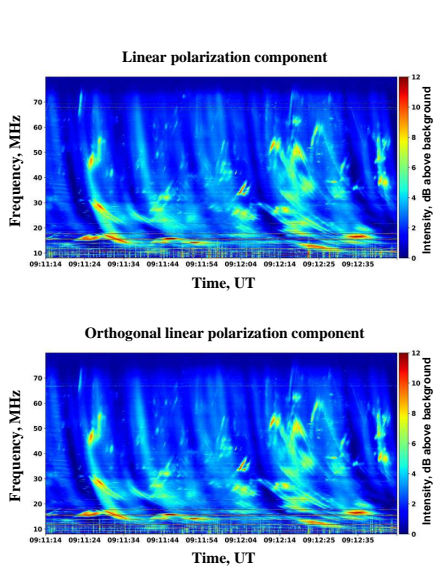}
}  \caption{The dynamic spectra show many DPs in two orthogonal linear polarizations. Many horizontal bright lines below 20 MHz and above 65 MHz are radio interferences. The records were obtained with the GURT antenna array from solar observations on July 12, 2017.} \label{gurt_sp}
\end{figure}

\begin{figure}
\centering
\resizebox{1.\columnwidth}{!}{%
  \includegraphics{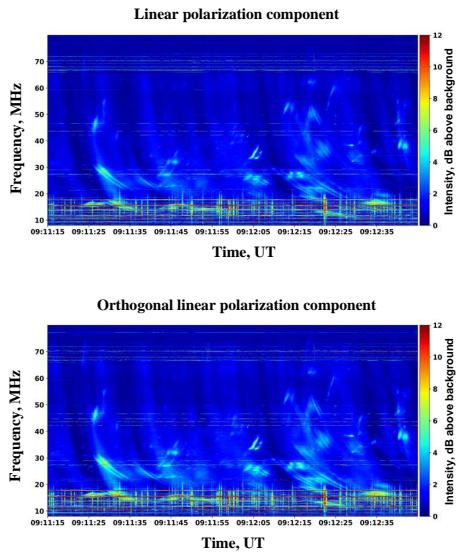}
}  \caption{The dynamic spectra were received with one crossed dipole. The measurements were carried out in two orthogonal linear polarizations at the same time as those shown in the Figure~\ref{gurt_sp}. Recall that each GURT subarray just consists of such dipoles. The sensitivity of any dipole is noticeably less than of a subarray GURT. That is why the number of DPs, detected by one dipole, was less.} \label{dipole_sp}
\end{figure}

\begin{figure}
\centering
\resizebox{1.\columnwidth}{!}{%
  \includegraphics{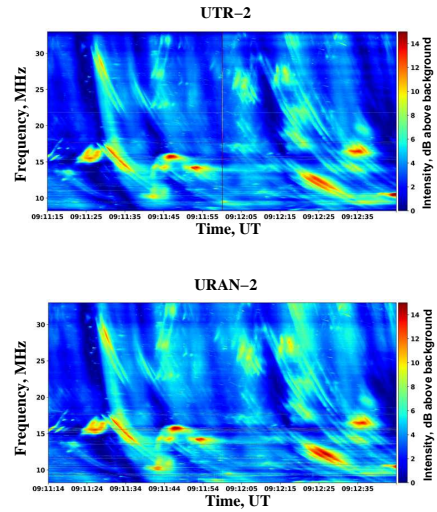}
}  \caption{The dynamic spectra were obtained with UTR--2 (top panel) and URAN--2 (bottom panel) arrays at the same time with GURT records (Figures~\ref{gurt_sp} and \ref{dipole_sp}).} \label{utr2+uran2_sp}
\end{figure}

In this observational session the solar radio emission was received with UTR--2, URAN--2, one of GURT subarrays as well as with a single crossed dipole antenna by the digital receiver/spectrometers DSPZ and ADR, respectively. They operated in the corresponding frequency ranges (8--32 MHz and 8--80 MHz, respectively) with the temporal resolution of 100 ms and the high frequency resolution. We have identified many DP bursts in which were forward and reverse ones. Their occurrence in frequency-time plane (dynamic spectrum) has a random nature. Examples of observed DPs are presented in the dynamic spectra shown in Figures~\ref{gurt_sp}, \ref{dipole_sp} and \ref{utr2+uran2_sp}. Although the bursts were similar in the appearance in dynamic spectra, they differed in the frequency bandwidth, starting and ending frequencies, total duration, incline and others. Surprisingly, despite their relatively low intensity, the bursts were observed even with a single crossed dipole antenna. This points to its good quality for radio astronomy observations. It should be noticed that the DPs were almost indistinguishable in the Nan\c{c}ay Decameter Array data (most likely because of insufficient time resolution) and were not visible at all in e--Callisto data (by reason of weak sensitivity) for the observations at frequencies below 100 MHz.

\begin{figure}
\centering
\resizebox{1.\columnwidth}{!}{%
  \includegraphics{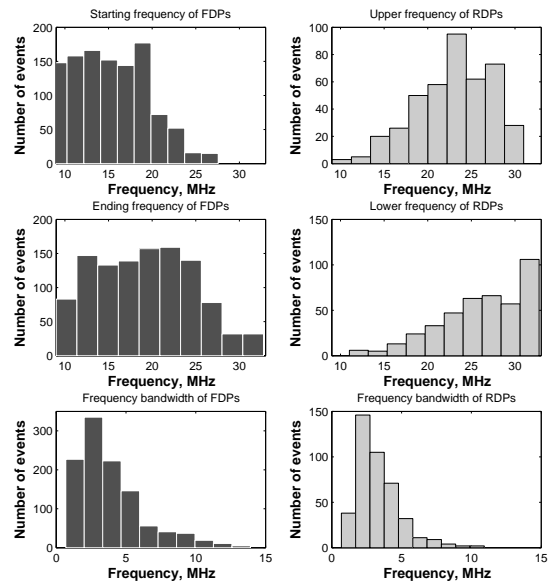}
}  \caption{Histograms describing statistical properties of FDPs and RDPs obtained in the solar observations with the UTR--2 on 12 July, 2017.} \label{hist-1}
\end{figure}

\section{Emergence of DPs below the ionosphere cutoff}\label{sectionempev}

The key problem of DP observations with the lunar farside surface is to find the lowest border of the frequency at which the bursts could be observed. As is well known (e.g., M{\o}ller-Pedersen {{et al.}} 1978), DPs arise during the type III burst storm. Generally the type III bursts are observed from some gigahertz up to tens kilohertz. This permits us to assume that DPs might occur in frequencies below the Earth's ionosphere cutoff. In favor of this remark the statistical analysis of the frequencies, at which each FDP (and RDP) begins and ends, is useful. For this reason, such a study has been fulfilled (Stanislavsky et al. 2017b), and it indicates that FDPs may be observed below 10 MHz. The study concerned to DPs on 10--12 July of 2015. To check this important result in new observations more subtly, we have considered our sample of DPs observed on 12 July of 2017. The observations were carried out with help of four sections of the UTR--2 by using DSPZ with time resolution 100 ms and frequency resolution 4 kHz.

The obtained histograms are shown in Figure~\ref{hist-1}. As can be seen, within the frequency range of UTR--2 observations the number of FDPs decreases at high frequencies, whereas the number of RDPs is reduced to lower frequencies. We try to find their distributions over frequency using the statistical approach from Stanislavsky et al. (2017b). The data included 760 DP bursts from which 550 were forward, and 210 were reverse. According to our consideration, the average value of frequency bandwidth was 3.31 $\pm$ 1.58 MHz for RDPs, whereas for FDPs it was 3.94 $\pm$ 2.38 MHz. There is a high correlation between starting and ending frequencies in DPs (0.92 for FDPs and 0.95 for RDPs). This indicates that starting and ending frequencies are related. In this connection it should be mentioned that the data contains outliers. Their appearance relates to the fact that some DPs were cropped in frequency above and/or below because of the instrumental capability to observe at 8--32 MHz. Such samples are censored (Cohen 1991). On the other hand, a part of DPs were unobservable because they in full occurred outside the frequency band of UTR--2, and they turn out to be truncated. This entangles the analysis of statistical properties of DPs. Nevertheless, there are good reasons to believe that the random variables $f_{1F}$ and $f_{2F}$ (as well as another pair $f_{1R}$ and $f_{2R}$) have a similar probabilistic distribution function (pdf). Using the robust fit less outlier--prone than the least--squares fit, we have the following relations
\begin{eqnarray}
f_{2F} & = & 1.19\,f_{1F}+e_F\,, \label{eq1}\\  
f_{2R} & = & 1.09\,f_{1R}+e_R\,, \label{eq2}    
\end{eqnarray}
where $e_F$ and $e_R$ are random noise, $f_1$ is the lower frequency, $f_2$ the upper one, and the additional subscripts $F$ and $R$ denote each set of the solar bursts (FDPs and RDPs, respectively).

The histograms of FDPs and RDPs presented in Figure~\ref{hist-1} indicate right--skewed data, described by a right--skewed distribution with a long right tail in the positive direction. The mean is also to the right of the peak. The skewness is real and is not caused with instrumental features. Note that such behavior is typical, if a random variable cannot take values less than zero. Most popular models used for non--negative data include the gamma, lognormal, and Weibull distributions (Feller 1966). To select the best fitting distribution for the histograms of Figure~\ref{hist-1}, we measured the ``distance'' between the data and the above--mentioned distributions by the goodness of fit test statistics, based on the Kolmogorov--Smirnov, Anderson--Darling, and Chi--Square tests (e.g., Stephens \& D'Agostino 1986). The gamma distribution model proves to describe the starting and ending frequencies of DPs in the best way.

Accounting for the relations (\ref{eq1}) and (\ref{eq2}), it is preferable to use the 3--parameter gamma (Pearson III type) distribution  (see Johnson et al. 1994). The distribution, also called the shifted gamma distribution, is written as
\begin{equation}
g(y) = \frac{1}{b\Gamma(a)}\left(\frac{y-m}{b}\right)^{a-1} \exp\left[-\frac{y-m}{b}\right]{\mathbbm{1}}_{(m,\infty)}\,,\label{eq3}
\end{equation}
where $a>0$ is the shape parameter, $b>0$ the scale parameter, $m$ the location (or shift) parameter. Note that the change $(Y-m)/b$ transforms the random variable $Y$ into the random one under the 2--parameter gamma pdf with the scale parameter equal to one. Using these sequences $(y_1,\dots, y_n)$, our next objective is to estimate the parameters $(a,b,m)$ of this distribution for FDPs and RDPs, respectively.

\begin{figure}
\centering
\resizebox{1.\columnwidth}{!}{%
  \includegraphics{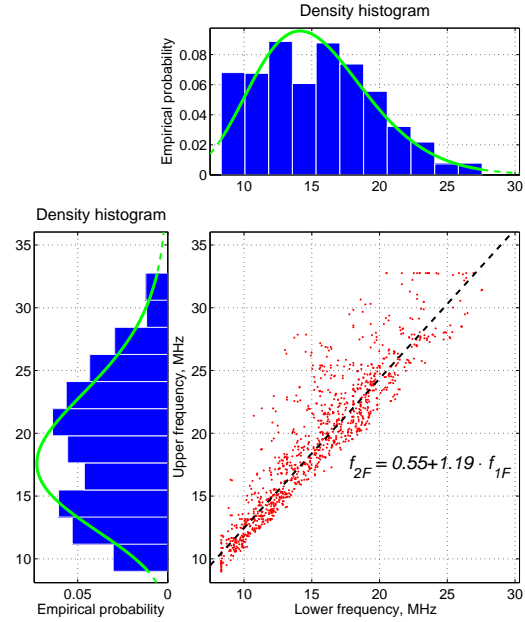}
}  \caption{Scatter plot with univariate normalized histograms for upper and lower frequencies of FDPs.} \label{hist_scatter}
\end{figure}

The method of moments is one of the simplest approaches to estimating the parameters. In this case it suffices to find the mean, variance and skewness (e.g., Grigoriu 2006). Then the pdf parameters of FDPs are $a_{2F}\approx11.84$, $b_{2F}\approx 1.63$ MHz, $m_{2F}=0$ MHz and $a_{1F}\approx 12.81$, $b_{1F}\approx 1.2$ MHz, $m_{1F}=0$ MHz for starting and ending frequencies, respectively. For RDPs we obtain $a_{1R}\approx17.51$, $b_{1R}\approx1.05$ MHz and $m_{1R}\approx4.62$ MHz for lower frequencies, whereas for upper frequencies they are $a_{2R}\approx11.05$, $b_{2R}\approx1.51$ MHz and $m_{2R}\approx9.6$ MHz. Fitting the FDPs histograms of Figure~\ref{hist-1} by the gamma distribution, it is easy to see (Figure~\ref{hist_scatter}) that there is the probability of observing the bursts below the ionosphere cutoff. The same may be done for RDPs. In particular, they will be visible above 32 MHz, and this is confirmed by direct observations with the GURT array. By a simple calculation of the number of DPs observed on July 12 with GURT records within 30--70 MHz (lower and higher frequencies were clogged by interference) we detected 2602 DPs from which 2178 events were RDPs, and 424 ones were FDPs. It should pointed out that the UTR--2 observations were carried out $\pm$ 3 hours before and after noon, whereas the GURT antenna allowed us to observe solar radiation longer, from dawn to dusk. In particular, this feature explains the difference in the number of DPs observed with the latter instrument.

\begin{figure}
\centering
\resizebox{1.\columnwidth}{!}{%
  \includegraphics{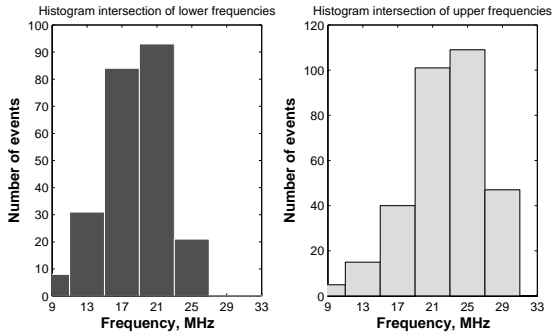}
}  \caption{Intersections of the histograms presented in Figure~\ref{hist-1}.} \label{intersection}
\end{figure}

According to the location of FDPs and RDPs in frequency, the occurrence of FDPs is more preferable at lower frequencies in comparison with RDPs. Besides, there is the low--frequency range where the emergence of RDPs and FDPs is almost equiprobable. Above these frequencies the reverse DPs prevail, whereas below the forward DPs dominate. The intersection peak (at 20--25 MHz on 12 July of 2017), corresponding approximately to the same amount of FDPs and RDPs, is shown in Figure~\ref{intersection}. Note that these conclusions are in good agreement with the results obtained in 2015 (Stanislavsky et al. 2017b).

\begin{figure}
\centering
\resizebox{1.\columnwidth}{!}{%
  \includegraphics{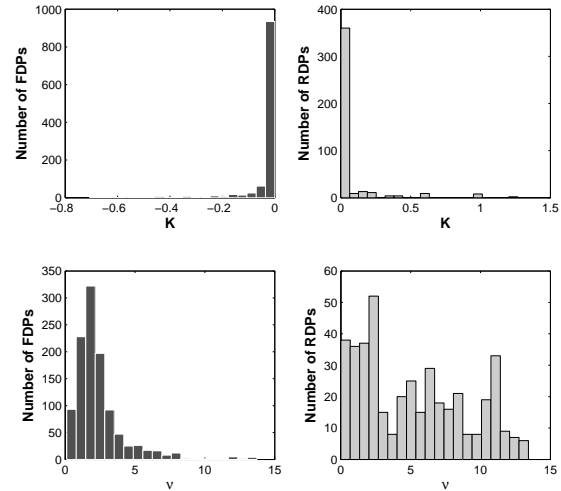}
}  \caption{Histograms of the parameters $K$ and $\nu$ from the equation (\ref{eq4}). The sign of $K$ indicates whether the frequency drift rate is positive (reverse DPs) or negative (forward DPs).} \label{K+nu}
\end{figure}

Using the spectral UTR--2 data, we briefly considered the frequency drift rates of DPs on July 12 by fitting the peak evolution of these bursts in frequency--time plane. Just as in the case of Stanislavsky et al. (2017a), they follow the power function
\begin{equation}
\frac{df}{dt} = Kf^\nu\,,\label{eq4}
\end{equation}
where $K$  and $\nu$  are the constants. Each DP component is characterized by a pair of the values. These parameters have very close values for components within any individual DP, but they are different from one DP to another. The histograms of their $K$ and $\nu$ are shown in Figure~\ref{K+nu}. In accordance with the histograms, the values $K$ and $\nu$ have characteristically skewed distributions. Similar results were obtained in radio observations of Stanislavsky et al. (2017b). For comparison, Alvarez and Haddock (1973) obtained the equation of frequency drift rates in the form $df/dt = - (0.01 \pm  0.008)\,f^{1.83 \pm 0.39}$ for many solar type III bursts within 75 kHz to 550 MHz. This means that the power dependence (\ref{eq4}) is not only characteristic of solar drift pairs, but also of other types of solar bursts.

\section{Discussion}\label{discussion}

The type II, III and IV solar radio bursts are currently observed below 10 MHz by satellite experiments above the terrestrial plasmasphere (Bougeret et al. 2008, Eastwood et al. 2010, Gopalswamy 2004, 2016; Kellogg 1980, Stone et al. 1992 and references therein). The success of these measurements is due to the fact that often the bursts are very intensive and sufficiently long in time. As many ground--based observations show, the diversity of solar radio bursts at low frequencies is much more, and they are various in flux and frequency--time characteristics. Unfortunately, the spacecraft capabilities do not give such high resolution in time and frequency as ground--based instruments. Besides, the sensitivity of low--frequency spacecraft antennas is not high yet because of their simple design (Stanislavsky et al. 2009). Therefore, observations of solar bursts with a radio telescope landed on the lunar farside at ultra--long wavelengths are so interesting.

Antenna is a heart of each modern radio telescope, and any good antenna is expensive, especially on the Moon (certainly with today’s technology). Ideally, such an antenna would have to be an array. Many ground--based radio arrays were preceded by prototypes consisting of a small number of antennas that would be scientifically productive. Therefore, the radio telescope based on the single crossed active antenna considered above could serve a starting point for a lunar radio astronomy observatory. Our measurements on the ground have shown that the GURT-type antenna is very effective instrument for radio astronomy purposes at low frequencies. If one adjusts its size, the dipole antenna will work at frequencies below 10 MHz without loss of quality. Such an antenna permits observing DPs, S bursts, absorption bursts, U bursts, zebra structures, J bursts and others. In addition, this does not exclude studies of more general problems: low--frequency sky mapping and monitoring, detection of pulsars, pathfinder measurements of the red--shifted HI line, emissions from the giant planets, etc. With help of the GURT antenna element at low frequencies we have already managed to observe the secular decrease of the flux density of Cassiopeia A and to detect radio emission from the pulsar B0809+74. Their detailed consideration goes beyond the scope of this work.

The GURT antenna element for two polarizations consists of a hollow metal tubes structure and antenna amplifier which makes it much easier to set up and transport. It is possible to develop an umbrella--like structure of telescopic tubes to unfold it on the Moon after transportation in built--up condition. The antenna amplifier can be developed in integrated circuit version that reduces its weight and power consumption. The consumption of the amplifier used in the GURT radio telescope is 180 mA $\times$ 12 V, but this can be easily reduced to 60 mA and even less with reducing a dynamic range. Accounting for quiet radio environment on the Moon, the crossed active dipole with ADC in 10--12 bits is sufficient to avoid most of the intermodulation leading to ADC overflow. The digital waveforms from the 2 polarized outputs of the crossed antenna at the frequency range with 20 MHz and below gives the data flow of about 800--960 Mbit/s. The transmission of such a signal from the lunar radio observatory to the Earth's station of space communications via a lunar satellite is quite realistic. The signal flow could be reduced with long accumulation of the radio signal or by performing on--line FFT like in DSPZ and ADR, if this is necessary.

Another important achievement of our solar observations is clear evidence of GURT capabilities in the study of DPs. This instrument is more than 2.5 times broadband than the UTR--2 radio telescope. Its frequency response is almost uniform from 10 to 80 MHz under high frequency--time resolution. Consisting of many crossed dipoles, the array allows measurements of the polarization characteristics, that is impossible for UTR--2. The GURT records cover almost completely the frequency range, where RDPs arise, as well as partially the frequencies typical for FDPs. In this paper, we do not aim to carry out a full analysis of DP properties from our observations. This will be done elsewhere.

Most of the astrophysical interpretations explained only a part of properties representative for DPs (e.g., Roberts 1958, Zheleznyakov 1965, M{\o}ller--Pedersen et al. 1978, Melrose 1982). The diversity of features with which the DPs occur in dynamic radio spectra indicates that the plasma processes take place in the upper corona. The random slopes of DPs in dynamic spectra, specially the vertical traces specify that the drift is not caused by the physical movement of any agency like a shock wave or an electron beam (Thejappa 1988). Therefore, one can assume a radiation mechanism that accounts for the observed drift rates without involving the movement of the exciting agency. According to the generation mechanism of Zaitsev \& Levin (1978), the frequency drift rate can be attributed to the shift of the resonance region rather than any physical agency. In particular, the theory helps to explain the occurrence of the DP chains and the vertical DP bursts recorded in solar observations. The mechanism, leading to the generation of DPs, suggests that the same stream of fast electrons, responsible for the type III bursts, might be also an origin for the generation of DPs. As the plasma emission is assumed to be the dominant emission mechanism for most radio bursts, DPs probably derive their energy from plasma waves created by many electron beams giving rise the solar storms of type III bursts. In particular, FDPs show the dependence of a drift rate on frequency similar to that of Type III radio bursts and S bursts (Stanislavsky et al. 2017a). A similar dependence holds for RDPs, but with a positive frequency drift.

Many common features (frequency bandwidth, central frequency and others) between two different sets of DP data, obtained in 2002 and 2015 from UTR--2 observations, point at the stability of the frequency--time properties of decameter DPs from one cycle of solar activity to another (Volvach et al. 2016). Having analyzed frequency--time properties of long DPs (Stanislavsky et al. 2017a) with the frequency bandwidth 8--15 MHz (there were about 7\% in the observational session of 2015), we have noticed that the best fitting of their frequency drift rate with frequency is very close to the drift rate dependence of S bursts in the same frequency range of observations (McConnell 1982); and recent studies of Morosan et al. 2015). In addition, the latter observations have shown indirectly  that the radio source of S bursts was located on the top of trans--equatorial loops connecting a complex active region in the southern hemisphere and a large area of bipolar plage in the northern hemisphere. Unfortunately, no decameter heliographic observations of the events, considered in this work, were made. Nevertheless, it is possible that DP bursts are a form of plasma emission, trapped by coronal magnetic loops connecting active regions that belonged to either $\beta$ or $\beta$--$\gamma$ magnetic classes, having both positive and negative magnetic polarities. Based on the generation mechanism of Zaitsev \& Levin (1978), the resonance layer, responsible for the DP radiation, is generated at different instant of time so that the DPs occurrence on the dynamic spectrum shows different slopes. If one assumes considerable fluctuations in the macroscopic parameters of solar corona such as electron density and magnetic field strength, then the frequency drifts are possible at all directions adjusted for the magnetic field configuration in coronal magnetic loops, and the histograms of DP occurrence over the frequency range characterizes the medium in which the fluctuations arise.

\section{Conclusions}
Thus, we have shown that solar DPs can be detected not only by antenna arrays, but even by the single crossed active dipole such as a component of the GURT array. This can be useful in the implementation of low--frequency radio telescopes on the Moon. The crossed active antenna works well in terrestrial conditions, even near ionospheric cutoff frequencies. There is full confidence that this tool allows significant radio astronomical observations from the lunar farside at very low frequencies, if this project is implemented. Howbeit, a subarray of the GURT is effective for studies of DPs from the ground. Therefore, such subarrays working together are able to surpass vastly any UTR--2 arm, used conditionally in solar observations, by its capabilities.

Our statistical analysis of upper and lower frequencies, at which DPs occur, clearly indicates that statistical properties of FDPs and RDPs are similar, and their pdf peaks are shifted in frequency. This explains why the most RDPs are detected at upper frequencies whereas FDPs prefer lower frequencies. These results allow us to assume with full confidence the possibility of observing FDPs below 10 MHz. In the future we are going to accomplish our detailed analysis of DP properties registered with UTR--2, GURT and URAN--2, adding information about frequency--time and polarization properties of DPs (Brazhenko et al. 2010). \\

We want to thank the GOES, STEREO, NDA and SOHO teams for developing and operating the instruments as well as for their open data policy. This research was partially supported by Research Grants 0117U002396 and 0118U000564 from the National Academy of Sciences of Ukraine.

\end{document}